# Origin of third order exceptional singularities and its signature in successive state conversion

Sayan Bhattacherjee, Arnab Laha, Somnath Ghosh*

Department of Physics, Indian Institute of Technology Jodhpur, Rajasthan-342011, India
**somiit@rediffmail.com*

**We report an open three-state perturbed system with quasi-statically varying Hamiltonian depending on the topological parameters. The effective system hosts two second order exceptional points (EP2s). Here a third order exceptional point (EP3) is explored with simultaneous encirclement of two EP2s by adiabatic variation of topological parameters. We study the robust successive state-exchange around the EP3. Applying adiabatic theorem, we estimate the evolution of total phase accumulated by each state during encirclement; where interestingly, the state-common to the pairs of coupled state picks up three times phase shift of 2π. Such an exclusively reported scheme can be exploited in potential applications of exceptional points, manipulating fewer topological parameters in various non-Hermitian systems.**

The unconventional features of open quantum-inspired/ wave-based systems, that interact with the environment, present a great deal in contemporary research field of various physical domain, especially in the field of photonics. These open systems are substantially described by the non-Hermitian Hamiltonians [1] where non-hermiticity arises in the form of gain and/or loss, etc. The distinctive characteristics of non-Hermitian systems and their key differences from Hermitian systems are now well established. A most common dynamical Hermitian feature, i.e., time dependent perturbation theory is no longer applicable for non-Hermitian formulations due to non-orthogonality of complex eigenstates [1]; where the solution of stationary Schrödinger equation should be possible, avoiding tedious time-dependent calculations. One of the exotic non-Hermitian feature is the existence of hidden singular points, coined as exceptional points (EP) by Kato [2], where the coupled eigenvalues are connected by a branch point. An EP is a particular point in the 2D (at least) parameter spectrum where the eigenvalues as well as the eigenfunctions coalesce [3, 4] and the underlying Hamiltonian becomes defective, i.e., it lacks a full basis of multivalued complex eigenvectors [4]. For example, consider a 2 × 2 non-Hermitian matrix [2]

$$M(\eta) = \begin{pmatrix} 1 & \eta \\ \eta^* & -1 \end{pmatrix}. \tag{1}$$

This yields the eigenvalues $E_{1,2} = \pm\sqrt{1 + |\eta|^2}$ and the eigenvectors $v_{1,2} = \begin{pmatrix} -\eta \\ 1 \mp \sqrt{1 + |\eta|^2} \end{pmatrix}$. They coalesce, i.e., $E_1 = E_2$ at a second order EP (EP2) in complex $\eta$-plane; where $\eta = \pm i$. After coalescence, the eigenstates lose their identities and pick up a huge magnitude. Note that, this coalescence phenomenon is far away from the conventional Hermitian degeneracies, usually occur at diabolic points.

Intense efforts have been put forward to explore the enticing features of EP2s in various non-Hermitian phenomena like unidirectional light transmission [5], manipulating the lasing-absorbing modes [6], asymmetric mode conversion/switching [7, 8], flip-of-states [9, 0], extreme enhancement in optical sensing [11, 12], etc. to name a few. However, instead of two levels coalescence at EP2s, lately, there are extensive attentiveness towards three levels coalescence with realization of a third order EP (EP3) [13–15]. In a system having more than two coupled eigenstates, one can identify multiple EP2s with proper parametric manipulation. Here an EP3 is realized with combined effect of at least two EP2s; where three coupled states are connected by a cube root branch point and analogically coalesced [13]. The fascinating properties of an EP3 are theoretically studied in different open systems like optical microcavities [15], waveguides [16], photonic crystals [17], atomic systems [10], etc. and experimentally explored in an acoustic cavity [18]. While an open system hosting an EP, the presence of non-hermiticity notably affects its performance; where the parameter dependent external perturbation gives rise miscellaneous exotic behaviors [8–10]. However, what happened if an EP3 is encountered in a system which itself changes adiabatically? If this can be implemented in a controlled manner, then it can be shown that with very little or negligible external effect one can observe certain exotic features of EPs; where for example EP aided-sensing can be done with extreme efficiency exploiting EP3 [19] in comparison with an EP2 [11]. In case of slowly varying Hamiltonian, perturbation correction [20] necessitate transitions between eigenstates due to presence of singularities with enormous increase in state amplitudes [2, 21]; which results breakdown in adiabatic theorem [21–23]. For such a higher order (more than two) system, multiple EP2s can be manipulated by parameter dependent internal/external perturbation; however, the system able to encounter an EP3 itself while changing adiabatically around at least two embedded EP2s along a closed parametric contour. This scheme is yet to be explored in the context of EP3.

In this letter, we report a three-state open system having quasi-statically varying Hamiltonian which is also subjected by a parameter dependent internal perturbation and study the non-Hermitian state-dynamics alongside an EP3. Here, the parameters are chosen and optimized in such a way that only one state must have to couple with the rest of two states and analytically connect with them via two EP2s. However, to explore evidence of an EP3, the interaction between two EP2s is controlled by varying the Hamiltonian itself very slowly within adiabatic limit, instead of back coupling between rest of the two uncoupled states. Such scheme is reported for the first-time to the best of our knowledge. The direct advantage of this scheme is that it requires less number of controlling parameters which would be more convenient for "proof of concept" of unconventional state-manipulation technique. The astonishing feature of our designed system is that due to adiabatic variation of controlled parameters along a closed contour around two EP2s, we exclusively observe EP3-driven state exchange phenomenon, a hallmark of the presence of an EP, among all the interacting states. We have also studied the total accumulated phase evolution of the system during encirclement; where state-common to the pair of coupled states picks up three times $2\pi$ phase, as a signature of an EP3. With proper optimization and parametric dependence, one may analogically realize the interactions between the quantum states associated with various real open systems, especially in optical domain.

In order to achieve our aspirations, we consider a 3 × 3-time dependent non-Hermitian Hamiltonian having the form $H(t) = H_0 + \lambda H_p$. Here three states are coupled in successive pairs as

$$H(t) = \begin{pmatrix} \tilde{\varepsilon}_1 & 0 & 0 \\ 0 & \tilde{\varepsilon}_2 & 0 \\ 0 & 0 & \tilde{\varepsilon}_3 \end{pmatrix} + \lambda \begin{pmatrix} 0 & \kappa_{12} & 0 \\ \kappa_{21} & 0 & \kappa_{23} \\ 0 & \kappa_{32} & 0 \end{pmatrix} \qquad (2)$$

In the passive part of the system $H_0$, $\tilde{\varepsilon}_j$ ($j = 1,2,3$) are the passive eigenvalues; where $\tilde{\varepsilon}_j = \varepsilon_j + \delta_j$. Here, $\delta_j$ represent three tunable complex parameters (as $\delta_j = \delta_{jR} + i\delta_{jI}$); where $\delta_{jR}$ control the adiabatic variation of $H_0$ over the decay rates $\delta_{jI}$ associated with $\varepsilon_j$. $H_0$ is subjected by an internal perturbation $H_p$ that depends on an independent complex parameter $\lambda$ (as, $\lambda = \lambda_R + i\lambda_I$). The elements of $H_p$ are customized as $\kappa_{12} = (i + \tilde{\varepsilon}_1 - \tilde{\varepsilon}_2 - \tilde{\varepsilon}_3)$, $\kappa_{21} = (\tilde{\varepsilon}_2 - \tilde{\varepsilon}_3)$, $\kappa_{23} = (\tilde{\varepsilon}_2 + \tilde{\varepsilon}_3)$ and $\kappa_{32} = 0.5\Re(\tilde{\varepsilon}_1 - \tilde{\varepsilon}_3)$. During operation, we numerically optimized the real passive eigenvalues as $\varepsilon_1 = 0.7$, $\varepsilon_2 = 0.65$ and $\varepsilon_3 = 0.3$. Associated decay rates $\delta_{jI}$ ($j = 1,2,3$) are fixed at 0.25. Moreover, an additional constraint $\delta_{2R} = 0 = \delta_{3R}$ is deliberately imposed. Thus, interactions between the eigenvalues $E_j$ ($j = 1,2,3$) of effective Hamiltonian H are controlled by tuning the complex $\delta_1$ over the independent parameter $\lambda$ within adiabatic limit. In the strong coupling limit, the individual states associated with pairs $(\tilde{\varepsilon}_1, \tilde{\varepsilon}_2)$ and $(\tilde{\varepsilon}_2, \tilde{\varepsilon}_3)$ should be interacted and may be analytically connected via two EP2s. However, in our proposed slowly changing Hamiltonian, the adiabatic tunability of $\delta_1$ itself introduces to couple $\tilde{\varepsilon}_1$ with $\tilde{\varepsilon}_3$, even in absence of proper backcoupling parameters in $H_p$ (as $\kappa_{13} = 0 = \kappa_{31}$).

In this context, the adiabatic theorem can be understood by considering two neighboring states separated by a finite gap, say $\Delta$, such that the system can realize the presence of a singularity via a transition of a finite excitation $\Delta$ between the corresponding states. To confirm the possibility of these transitions, one can consider the Hamiltonian $H(\omega)$ in frequency domain via Fourier transform of $H(t)$. If the time dependence of $H$ is made sufficiently slow, $H(\omega)$ will only have finite matrix elements for $\omega \ll \Delta$. As a result, the system will remain in its instantaneous eigenstate. This phenomenon is referred as the adiabatic evolution. It can be proved by the help of work done by Born and Fock in 1928 [24]. If $|n(t)\rangle$ is an orthonormal set of instantaneous eigenstates of $H(t)$ with eigenvalues $E_n(t)_n$, then the exact solution of Schrödinger equation can be expressed as:

$$|\Psi(t)\rangle = \sum_n c_n(t)|n(t)\rangle e^{-i\phi_D^n(t)}. \tag{3}$$

Thus, the solutions of the Hamiltonian can be written as the sum of orthonormal eigenstates $|n(t)\rangle_n$; where $|n(t)\rangle_n$ are associated with a phase $\phi_D^n(t) = \int_{t_0}^{t} E_n(\tau)d\tau$, known as dynamical phase. Here, the coefficient $c_n(t)$ can be derived from the Schrödinger equation under adiabatic condition; where the final forms can be provided as:

$$c_n(t) = c_n(t_0) \exp\left\{-\int_{t_0}^{t} \left\langle n \left| \frac{d}{dt} \right| n \right\rangle d\tau\right\}. \tag{4}$$

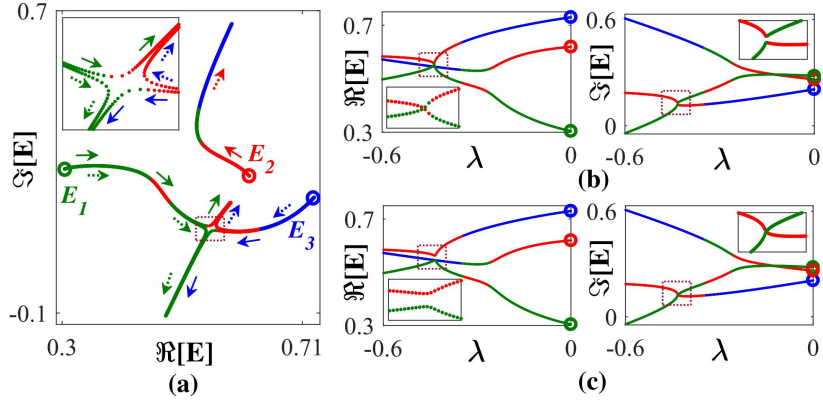

**Fig. 1. (a)** Trajectories of $E_j$ (designated by green, red and blue dots for $j = 1,2,3$ respectively) with $\lambda$; when $E_1$ and $E_3$ exhibit ARCs (unaffecting $E_2$) with corresponding **(b)** crossing and anti-crossing in $\Re[E]$ and $\Im[E]$ w.r.t. $\lambda_R$ for $\delta_{1R} = 0.003$; and **(c)** vice-versa for $\delta_{1R} = 0.004$ respectively. Circular markers with respective colors indicate the initial positions of $E_j$. In (a) the direction of evolutions are shown by the arrows with respective colors (dotted arrows are used for lower $\delta_{1R}$ whereas solid arrows for higher $\delta_{1R}$). The regions marked by brown dotted rectangles are zoomed in respective insets; where we deliberately omit the unaffected state for clear visibility.

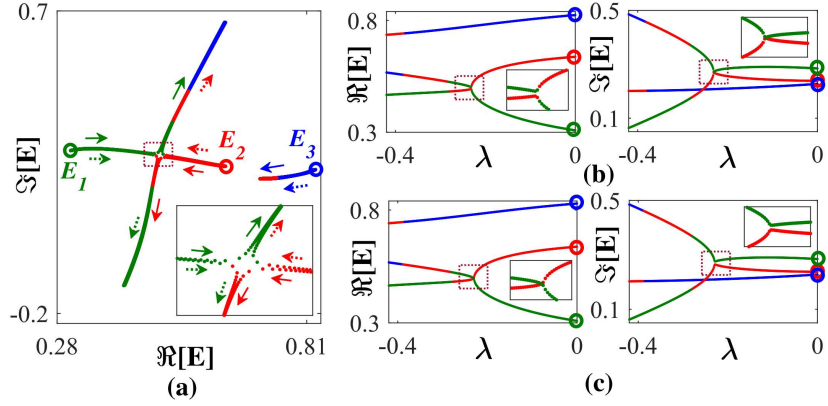

**Fig. 2. (a)** Similar trajectories of $E_j$; when $E_1$ and $E_2$ exhibit ARCs (unaffecting $E_3$) with corresponding **(b)** anti-crossing and crossing in $\Re[E]$ and $\Im[E]$ for $\delta_{1R} = 0.12$; and **(c)** vice-versa for $\delta_{1R} = 0.125$ respectively. The notations carry the same meaning as described for Fig. 1.

Now, under adiabatic changes in the proposed Hamiltonian $H$, the interaction between $E_1$ and $E_2$ as well as $E_1$ and $E_3$ are studied with implementation of their dynamical evolutions, exploiting the phenomena of special avoided resonance crossings (ARC) [4]. In Fig. 1 and 2 the dynamics of $E_j$ ($j = 1, 2, 3$) are plotted in the complex eigenvalue-plane ($E$-plane) with extremely slow complex variation of $\lambda$ in a specified range for different $\delta_{1R}$-values. Here, to achieve controlled interaction phenomena between $E_j$, with independent tunable range of $\lambda_R$, we restrict the simultaneous variation of $\lambda_I$ in the same tunable range with exact equal adiabaticity. Because, if we tune only $\lambda_R$ fixing $\lambda_I$, then $E_1$ is only able to interact with either $E_2$ or $E_3$, based on the choices of other parameters. In Fig. 1, the interactions between $E_1$ and $E_3$ are demonstrated via a special ARC in the complex eigenvalue plane ($E$-plane) with respect to $\lambda_R$ (while $\lambda_I$ simultaneously varies with $\lambda_R$). For $\delta_{1R} = 0.003$, they exhibit ARC in Fig. 1(a) (directed by dotted arrows of respective colors) with crossing in $\Re[E]$ and anti-crossing in $\Im[E]$ with respect to $\lambda_R$ as shown in Fig. 1(b). Now for slightly higher value of $\delta_{1R} = 0.004$, a different kind of ARC occurs along the direction shown by solid arrows of respective colors in Fig. 1(a); where $\Re[E]$ experiences anti-crossing with simultaneous crossing in $\Im[E]$ with respect to $\lambda_R$ as displayed in Fig. 1(c). However, for both $\delta_{1R}$ values $E_2$ moves along the same directions unaffecting the interactions between $E_1$ and $E_3$. Such instantaneous transitions for two different $\delta_{1R}$-values clearly indicate the presence of an EP2 in ($\lambda_R, \delta_{1R}$)-plane at ($-0.432, 0.0035$), say EP2[(1)]. Similarly, we observe similar heterogeneous behaviors in crossing/ anticrossing of $\Re[E]$ and $\Im[E]$ corresponding to the special ARCs between $E_1$ and $E_2$ (unaffecting $E_3$) (as shown in Figs. 2(a)) while

$\delta_{1R}$ is changed suddenly from 0.12 to 0.125; as shown in Fig. 2(b) and (c) respectively. This indicates the presence of another EP2 at ($\lambda_R = -0.23, \delta_{1R} = 0.1225$), say EP2$^{(2)}$.

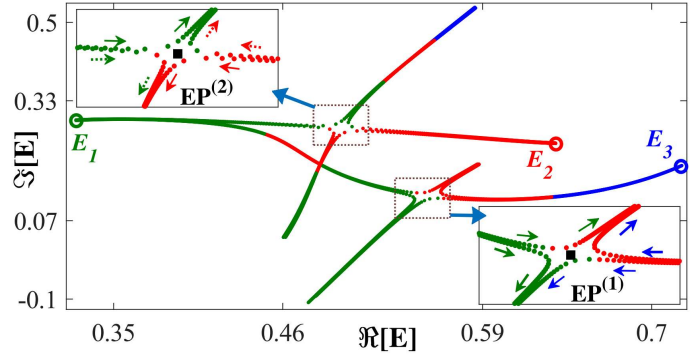

**Fig. 3.** Three-state-ARC associated with all the three interacting states is shown at a time; where in the zoomed portions the black square markers indicate the approximate equivalent positions of two EP2s in the complex E-plane.

Thus, we identify two EP2s in ($\lambda_R, \delta_{1R}$)-space; where $E_1$ is interacted with both $E_2$ and $E_3$ for two different parametric choices. However, both the EP2s are also associated with other two parameters viz. tunable $\lambda_I$ and fixed $\delta_{1I}$ (decay rate of $\tilde{\varepsilon}_1$). As $\lambda_I$ is simultaneously varied with $\lambda_R$ in same tunable range with exact equal adiabaticity, at EP2$^{(1)}$ $\lambda_I$ takes the value $-0.432$ whereas at EP2$^{(2)}$, $\lambda_I = -0.23$. Again, if we choose another $\delta_{1I}$ (except 0.25 as indicated in previous), the positions of two EP2s should be different. Thus, at both the EP2s, $\delta_{1I} = 0.25$. Interestingly, due to such simultaneous effects of other parameters, branch cuts (a singularity-assisted peculiar phenomena) [7] appears on the dynamics of $E_j$. The changes of colors (representing individual eigenvalues) as observed in the Fig. 1 and 2 indicate such branch cuts. Therefore, we behold two situations with respect to choice of $\delta_{1R}$ over the specified $\lambda$-span, where among three interacting states; any two states are interacting keeping the third as an observer. The whole phenomena can be referred as three-state-ARC which are presented in Fig. 3. In the both insets, the black square markers indicate the equivalent positions of two EP2s in complex $E$-plane. This is an exclusive signature which we explore identify an EP3 by combined effect of both the EP2s with the help variation of an additional parameter [13, 14]. Following an adiabatic parametric encirclement around both the EP2s simultaneously, one should observe successive conversion/switching (with exchanging their positions in eigenvalue-plane) [4, 8, 9] between the three coupled eigenvalues [10]; which is a deficient hallmark of effect of an EP3 [13, 14]. However, for our proposed Hamiltonian, as there are no parameters in $H_p$ for coupling between $\tilde{\varepsilon}_1$ and $\tilde{\varepsilon}_3$, it would be inevitable that in spite of having a complete loop in perturbation parameter space, the coupled states would not make a complete loop in eigenvalue plane after mutually exchanging their positions (not the general case [10]). Here, the most interesting part of the adiabatic evolution is associated with cyclic time dependence of $H$. The key feature implemented in this work is that we have varied the system adiabatically based on the complex tunable parameter $\delta_1$ only; where two EP2s are simultaneously encircled in $(\delta_{1R}, \delta_{1I})$-plane. Here the effect of an EP3 gives a complete state conversion as described in the preceding section.

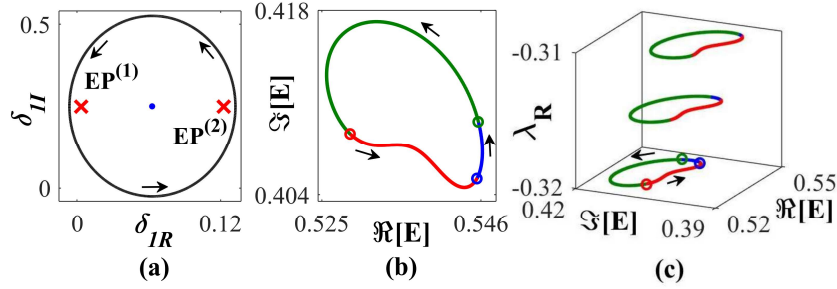

**Fig. 4. (a)** Parametric encirclement around two identified EP2s (shown by two red crosses) in ($\delta_{1R}, \delta_{1I}$)-plane with center at ($\delta_{1R}^c = 0.063, \delta_{1I}^c = 0.25$) and radius $a = 0.98$. **(b)** Complex trajectories of all three interacting eigen-values exhibiting successive state conversion/switching for an anti-clockwise encirclement along the closed loop described in (a) for a fixed $\lambda_R = -0.32$. **(c)** Robustness of flip-of-states phenomena as described in (b) with respect to $\lambda_R$. Arrows indicate the direction of progressions.

Accordingly, in the complex δ1-plane, we locate EP2$^{(1)}$ at $(0.0035, 0.25)$ and EP2$^{(2)}$ at $(0.1225, 0.25)$, and also consider a closed contour to enclose them properly as shown in Fig. 4(a) following the equation $\delta_1(\theta) = \delta_1^c + a(\delta_{1R}^c \cos\theta + i\delta_{1I}^c \sin\theta)$. Here, $\delta_1^c = (\delta_{1R}^c + \delta_{1I}^c)$ represent the centre and $a \in (0, 1]$ is a characteristics parameter. A tunable angle $\theta \in [0, 2\pi]$ control the adiabatic variation of the system via the complex $\delta_1$ around two EP2s. Interestingly, for one complete loop in parameter space, three interacting states are exchanging their positions successively for a fixed $\lambda_R = -0.32$ ($\lambda_I$ also takes same value). This successive state-conversion, namely flip-of-states phenomena is displayed in Fig. 4(b). In Fig. 4(c), we present the robustness of the described flip-of-states with respect to $\lambda_R$ with simultaneous variation in $\lambda_I$. Here as can be seen, the successive flip-of-states is omnipresent even in variation of $\lambda_R$. As the described flip of-states phenomena is independent with variation in $\lambda$; such robustness as described in Fig. 4(c) exhibits the topological protection of the state-conversion mechanism during $\lambda$-dependent perturbation correction.

Now the cyclic time dependence of $H$ is linked with the Berry phase [23] which can be understood from the Eq. 4; where the phase factor $-\int_{t0}^{t}\langle n|\frac{d}{dt}|n\rangle d\tau$ can't be gauged away commonly. Considering $R(t)$ as time dependence of $H$, the cyclic evolution can define a loop $\gamma: t \to R(t), t \in [0,T], R(0) = R(T)$ in the parameter space. Now, using the relation $\langle n|\frac{d}{dt}|n\rangle = \langle n|\partial_\mu|n\rangle \dot{R}^\mu$ ($\partial_\mu = \partial/\partial \dot{R}^\mu$) and exploiting Stokes theorem we can calculate the Berry phase ($\varphi_\gamma^B$) associated with the loop $\gamma$ as

$$-\oint_T \langle n|\frac{d}{dt}|n\rangle d\tau = \int_\gamma A^B = \int_S dA^B = \int_S F^B = \varphi_\gamma^B \tag{5}$$

Here, the Berry's curvature $F^B (= F_{\mu\nu}^B dR^\mu \wedge dR^\mu)$ can be defined as $F^B = -i[\langle \partial_\mu n|\partial_\nu n\rangle - \langle \partial_\nu n|\partial_\mu n\rangle] = 2\Im\langle \partial_\nu n|\partial_\mu n\rangle$. $F_B$ is a gauge invariant quantity which is analogous to the field strength tensor in electrodynamics. Thus Eq. 4 can be written as $c_n(T) = c_n(0) \exp(i\varphi_\gamma^B)$. Therefore, the Berry phase ($\varphi_\gamma^B$) is gauge invariant due to interference effects between coherent superposition that undergo different adiabatic evolutions. $\varphi_\gamma^B$ is a purely geometrical quantity which only depends on the inner-geometrical relation of the family of states $|n(R)\rangle$ along the loop $\gamma$. Thus contextually, during evolution of states $\langle n|$ to the state $|m\rangle$ followed by a parametric encirclement around the respective EP, the phase difference between the final and initial state associates with the variations of both the dynamical phase and the geometric phase.

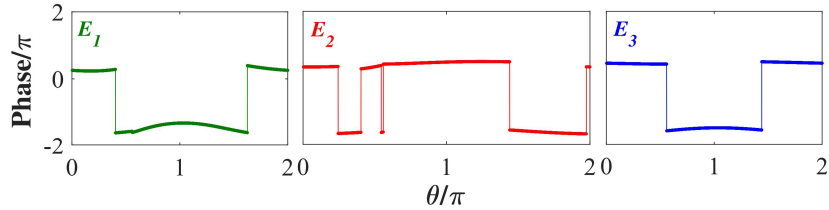

**Fig. 5.** Phase evolution of individual states along the closed contour described in Fig. 4(a) for a fixed $\lambda_R = -0.32$.

In Fig. 5, we display the total phase evolutions of three interacting states for an adiabatic change of H along a closed contour around two EP2s in complex $\delta_1$-plane. Estimating the evolutions of the eigenfunctions for a fixed $\lambda_R = -0.32$, we calculate the corresponding total accumulated phases at each point on the parametric loop (i.e. for each $\theta$ values) as described in Fig. 4(a). It is evident in Fig. 5 that the eigenstates correspond to $E_1$ and $E_3$ pick up $2\pi$ phase shift once; whereas the state corresponds ro $E_2$ experiences this $2\pi$ phase shift three times. Thus, this phase dynamics evidently exhibits the hallmark of both EP2s as well as one EP3. However, in contrary, for a double cyclic parametric rotation around a single EP2, the total geometric phase accumulated is equal to ±π. While, in case of two encircled EP2s, the total phase accumulated due to one EP2 is canceled by the accumulated phase of other one [15]. Thus, the total accumulated phase for one complete parametric rotation around two EP2s is either 0 or 2π.

In summary, a three state internally perturbed non-Hermitian Hamiltonian, hosting three interacting eigen-states is proposed and modeled. Here, the passive Hamiltonian is able to vary itself very slowly based on a chosen specific parameter, i.e. $\delta_1$. The perturbation term is customized in such a way that even in absence of proper back coupling parameters, all three states are mutually coupled in a successive way and analytically connected via two EP2s. Here an EP3 is explored by interaction between two identified EP2s; which is controlled by adiabatic variation of the system itself. We properly encircle two EP2s in complex $\delta_1$ plane and explore the presence of an EP3. With such parametric encirclement, the successive switching among three coupled consecutive states in complex eigenvalue plane is reported for the first time; where such flip-of-states phenomena is also robust against the variation in perturbation. We also report the phase variation of each states around EP3; where a clear signature of phase picks up of three times of 2π around such EP3 is observed during a complete encirclement. The systems realized with such exclusively proposed scheme may open up an extensive platform for a wide range of EP-aided state-of-the-art integrated applications like all-optical on-chip mode converters, circulators, filters, ultra-sensitive EP-aided optical sensors etc.

## ACKNOWLAGEMENT